%% file: HION-2017-10-PAPER.tex
\newcommand*{\ATLASLATEXPATH}{latex/}

\pdfoutput=1

\documentclass[cernpreprint, atlasdraft=true, texlive=2016, UKenglish]{\ATLASLATEXPATH atlasdoc}

\usepackage[biblatex=true,backend=bibtex]{\ATLASLATEXPATH atlaspackage}
\usepackage{\ATLASLATEXPATH atlasbiblatex}
\usepackage{\ATLASLATEXPATH atlasphysics}

\graphicspath{{logos/}{figures/}}

\usepackage{HION-2017-10-PAPER-defs}

\input{HION-2017-10-PAPER-metadata}

\hypersetup{pdftitle={ATLAS document},pdfauthor={The ATLAS Collaboration}}

\AtlasJournalRef{Phys. Lett. B 790 (2019) 108}
\AtlasDOI{10.1016/j.physletb.2018.10.076}

\begin{document}

\maketitle

\section{Introduction}
\label{sec:introduction}

  Heavy-ion collisions at ultra-relativistic energies produce a hot, 
dense medium of strongly interacting nuclear matter understood to be 
composed of unscreened colour charges which is commonly called a 
quark--gluon plasma (QGP)~\cite{Adcox:2004mh,Adams:2005dq,Back:2004je,Arsene:2004fa}. 
  Products of 
  the hard scattering of quarks and gluons occurring in these collisions 
  evolve as parton showers that propagate through the hot medium. 
  Parton shower constituents emit medium-induced gluon 
  radiation or suffer from elastic scattering processes
  and as a consequence they lose energy, leading to the formation of lower-energy jets. 
 This 
phenomenon is termed ``jet 
quenching''~\cite{Mehtar-Tani:2013pia,Qin:2015srf,Blaizot:2015lma}. 
It has been directly observed as the suppression of the jet yields in \PbPb\ collisions
compared to jet yields in \pp\ collisions~\cite{Aad:2012vca,Aad:2014bxa,Khachatryan:2016jfl,Abelev:2013kqa}, 
the modification of jet internal structure~\cite{Aaboud:2017bzv,Chatrchyan:2014ava,Aad:2014wha,Khachatryan:2016tfj}, 
and a significant modification of the transverse energy balance in 
dijet~\cite{Aaboud:2017eww,Aad:2010bu,Chatrchyan:2011sx} and multijet systems~\cite{Aad:2015bsa}.

  The energy loss of partons propagating through the QGP results in a reduction 
of the jet yield at a given transverse momentum (\pt). This together with the falling shape of the jet \pt\ spectrum lead to 
the observed suppression of jets in collisions of nuclei relative to 
 \pp\ collisions.
  Central heavy-ion collisions have an enhanced hard-scattering rate due to the larger geometric overlap between the colliding nuclei, resulting 
  in a larger per-collision nucleon--nucleon flux. 
  To quantitatively assess the quenching effects, the hard-scattering rates measured in \PbPb\ collisions are normalised by the mean nuclear 
thickness function, \TAA, which accounts for 
 this geometric enhancement~\cite{Miller:2007ri}.
  The magnitude of the inclusive jet suppression in nuclear collisions relative to \pp\ is quantified by the nuclear modification
factor

  \begin{equation} 
  \nonumber
  \RAA=\dfrac{ \left.\dfrac{1}{N_{\mathrm{evt}} }\dfrac{\diff^2 
N_{\mathrm{jet}}}{\diff \pt \diff y} \right|_{\mathrm{cent}}
}
{
\left. \TAA \dfrac{\diff^2 \sigma_{\mathrm{jet}}}{\diff \pt \diff y}\right|_{pp}
}\, ,
  \label{eqn:raa_def}
  \end{equation} 
  where $N_{\mathrm{jet}}$ and $\sigma_{\mathrm{jet}}$ are the jet yield in 
\PbPb\ collisions and the jet cross-section in \pp\ collisions, 
respectively, both measured as a function of transverse momentum, \pt, and rapidity, $y$, and where $N_{\mathrm{evt}}$ is the total number of 
  \PbPb\ collisions within a chosen centrality interval.

  A value of $\RAA \approx 0.5$ in central collisions was reported in 
\PbPb\ measurements at $\sqrtsnn = 2.76$~\TeV\ by the ATLAS and CMS 
Collaborations for jet \pt\ above 100~\GeV~\cite{Aad:2014bxa,Khachatryan:2016jfl}.
  These measurements therefore show a suppression of jet yields by a factor of 
two in central collisions relative to the corresponding \pp\ 
yields at the same centre-of-mass energy. Also a clear centrality dependence is observed.
  Two unexpected features \cite{Abreu:2007kv} also emerge from those studies: \RAA\ increases 
only very slowly with increasing jet \pt, and no dependence of \RAA\ on 
jet rapidity is observed. Measurements by the ATLAS and CMS Collaborations can be complemented by the measurement by the ALICE Collaboration which reports 
\RAA\ for jets measured in \pt\ interval of $30 - 120$~GeV in central \PbPb\ collisions \cite{Adam:2015ewa}.

  This Letter describes the 
  new measurements of yields of $R=0.4$ anti-$k_t$ 
jets~\cite{Cacciari:2008gp} performed with 0.49 nb$^{-1}$ of \PbPb\ data 
collected at $\sqrtsnn = 5.02$~\TeV\ in 2015 and 25~pb$^{-1}$ of \pp\ 
data collected at $\sqrts = 5.02$~\TeV\ in the same year.
  This new study closely follows the first measurement by 
the ATLAS Collaboration~\cite{Aad:2014bxa} performed using 
0.14 nb$^{-1}$ of \PbPb\ data collected at $\sqrtsnn = 2.76$~\TeV\ in 2011
and
4.0~pb$^{-1}$ of \pp\ data collected at $\sqrts = 2.76$~\TeV\ in 2013.
  Higher luminosity, increased centre-of-mass energy, and improved 
analysis techniques allowed to extend the measurement to more than two 
times higher transverse momenta, and to larger rapidities. This new 
measurement provides input relevant to a detailed theoretical 
description of jet suppression, especially its dependence on the 
collision energy, centrality, jet $\pt$, and rapidity.

\section{Experimental setup}
\label{sec:exp}

The ATLAS experiment~\cite{Aad:2008zzm} at the LHC features a multipurpose particle detector with a forward-backward
symmetric cylindrical geometry and a nearly full coverage in solid angle.\footnote{
  ATLAS uses a right-handed coordinate system with its origin at the 
nominal interaction point (IP) in the centre of the detector and the 
$z$-axis along the beam pipe. The $x$-axis points from the IP to the 
centre of the LHC ring, and the $y$-axis points upward. Cylindrical 
coordinates $(r,\phi)$ are used in the transverse plane, $\phi$ being 
the azimuthal angle around the beam pipe. The pseudorapidity is defined 
in terms of the polar angle $\theta$ as $\eta=-\ln\tan(\theta/2)$. 
Rapidity $y$ is defined as $y = 0.5 \ln [(E+p_z)/(E-p_z)]$ where $E$ 
and $p_z$ are the energy and the component of the momentum along the 
beam direction, respectively. Angular distance is measured in units of $\Delta R \equiv \sqrt{ (\Delta\eta)^2+(\Delta\phi)^2}$.}
The measurements presented here were performed using the ATLAS 
inner detector, calorimeter, trigger and data acquisition systems.

The inner-detector system (ID) is immersed in a 2~T axial magnetic field 
and provides charged-particle tracking in the pseudorapidity 
range $|\eta| < 2.5$. 
  The high-granularity silicon pixel detector covers the 
vertex region and typically provides four measurements per track.
It is followed by the silicon microstrip tracker (SCT) which comprises 
four cylindrical layers of double-sided silicon strip detectors in the 
barrel region, and nine disks in each endcap. These silicon detectors 
are complemented by the transition radiation tracker, a drift-tube-based detector,
which  surrounds the SCT and has coverage up to $|\eta| = 2.0$.

The calorimeter system consists of a sampling lead/liquid-argon 
(LAr) electromagnetic (EM) calorimeter covering $|\eta|<3.2$, a 
steel/scintillator sampling hadronic calorimeter covering $|\eta| 
<1.7$, a LAr hadronic calorimeter covering $1.5 < |\eta| < 3.2$, and two 
LAr forward calorimeters~(FCal) covering $3.1 < |\eta| < 4.9$. 
  The  hadronic  calorimeter  has  three  sampling  layers  longitudinal  in
shower depth in $|\eta| < 1.7$ 
and four sampling layers in $1.5 < |\eta| < 3.2$, with a slight overlap.
  The EM calorimeter is segmented longitudinally in shower depth into
three compartments with an additional pre-sampler layer.

  A two-level trigger system~\cite{Aaboud:2016leb} 
was used to select the \PbPb\ and \pp\ 
collisions analysed here. The first level (L1) is a hardware-based 
trigger stage which is implemented with custom electronics. The second 
level is the software-based high-level trigger (HLT). The events were 
selected by the HLT which was seeded by a L1 jet trigger, total energy trigger, or zero-degree calorimeter (ZDC) trigger.
The total energy trigger required a total transverse energy 
measured in the calorimeter system to be greater than 5~\GeV\ in \pp\ 
interactions and 50~\GeV\ in \PbPb\ interactions. 
  The ZDC trigger required a presence of at least one neutron on both 
sides of ZDC ($|\eta|>8.3$).
  The HLT jet 
trigger used a jet reconstruction algorithm 
similar to the \PbPb\ one applied in offline 
analyses. It selected events containing jets with transverse energies 
exceeding a threshold, using a range of thresholds up 
to~100~\GeV\ in \PbPb\ collisions and up to 85~\GeV\ in \pp\ collisions. 
In both the \pp\ and \PbPb\ collisions, the highest-threshold jet trigger 
sampled the full delivered luminosity while all lower threshold triggers were 
prescaled.

In addition to the jet trigger, two triggers were used in \PbPb\ 
collisions to select minimum-bias events. The minimum-bias trigger 
required either more than 50~\GeV\ transverse energy recorded in the 
whole calorimeter system by L1 trigger or a signal from the ZDC trigger and a track 
identified by the HLT.

\section{Data and Monte Carlo samples, and event selection}
\label{sec:data}

The impact of detector effects on the measurement was determined 
using a simulated detector response evaluated by running Monte Carlo 
(MC) samples through a \textsc{Geant4}-based detector simulation 
package~\cite{Agostinelli:2002hh,Aad:2010ah}. Two MC samples were used 
in this study. 
  In the first one, multi-jet processes were simulated with ~\textsc{Powheg-Box}~v2 \cite{Nason:2004rx,Frixione:2007vw,Alioli:2010xa} interfaced to 
the~\textsc{Pythia}~8.186 \cite{Sjostrand:2007gs,Sjostrand:2006za} parton shower model. The CT10 PDF set~\cite{Lai:2010vv} was used in the matrix element while the A14 
set of tuned parameters \cite{ATL-PHYS-PUB-2014-021} was used together with the NNPDF2.3LO PDF set \cite{Ball:2012cx} for the modelling of the non-perturbative effects. 
The EvtGen 1.2.0 program \cite{EvtGen} was used for the properties of b- and c-hadron decays.
  In total, 
$2.9 \times 10^7$ hard-scattering events at $\sqrt{s}=5.02$~\TeV\ were simulated at the NLO precision, spanning a range of jet transverse 
momenta from 20 to 1300~\GeV. The second MC sample consists 
of the same signal events as those used in the first sample but 
embedded into minimum-bias \PbPb\ data events. This minimum-bias sample was combined 
with the signal from \textsc{Powheg+Pythia}8 simulation at the digitisation 
stage, and then reconstructed as a combined event. 
  So-called ``truth jets'' are defined by applying the anti-$k_t$ algorithm with radius parameter $R=0.4$ to stable particles in the MC event generator's output, defined as those with a proper lifetime greater than 10~ps, but excluding muons and neutrinos, which do not leave significant energy deposits in the calorimeter.

The level of overall event activity or centrality in \PbPb\ 
collisions is characterised using the sum of the total transverse 
energy in the forward calorimeter, $\ETfcal$, at the electromagnetic energy 
scale. The \ETfcal\ distribution is divided into percentiles of the 
total inelastic cross-section for Pb+Pb collisions with 0--10\% centrality interval classifying the most central collisions.
  The minimum-bias 
trigger and event selection are estimated to sample $84.5\%$ of the total
inelastic cross-section, with an uncertainty of 1\%. 
  A Glauber model analysis of the \ETfcal\ distribution is used to evaluate $\TAA$ and the number of nucleons 
participating in the collision, $\ANpart$, in each centrality interval~\cite{Miller:2007ri,Loizides:2014vua,ATLAS:2016tor}. The centrality intervals used in this measurement are indicated in Table~\ref{tbl:npart} along with their respective calculations of $\ANpart$ and $\TAA$.

\begin{table}[h]
\caption{
  The mean number of participants, $\ANpart$, the mean nuclear thickness 
function, $\TAA$, and their uncertainties (see Section~\ref{sec:uncert}) for different centrality intervals.
  }
\centering
\begin{tabular}{|  r | r | r |} \hline
Centrality range &	\ANpart\ 	&\TAA\ [1/mb]  \\ \hline
70--80\%  	 &   	15.4 $\pm$ 1.0 &	0.22 $\pm$ 0.02  \\ \hline
60--70\%  	 &   	30.6 $\pm$ 1.6 &	0.57 $\pm$ 0.04    \\ \hline
50--60\%  	 &   	53.9 $\pm$ 1.9 &	1.27 $\pm$ 0.07  \\ \hline
40--50\%  	 &   	87.0 $\pm$ 2.3 &	2.63 $\pm$ 0.11  \\ \hline
30--40\%  	 &   	131.4 $\pm$ 2.6 &	4.94 $\pm$ 0.15  \\ \hline
20--30\%  	 &   	189.1 $\pm$ 2.7 &	8.63 $\pm$ 0.17  \\ \hline
10--20\%  	 &   	264.0 $\pm$ 2.8 &	14.33 $\pm$ 0.17  \\ \hline
0--10\%     	 &   	358.8 $\pm$ 2.3 &	23.35 $\pm$ 0.20 \\ \hline
\end{tabular}
\label{tbl:npart}
\end{table}

Jets used in this analysis are reconstructed either in minimum-bias 
events or in events selected by inclusive jet triggers in the region of jet \pt\ for which 
the trigger efficiencies are greater than 99\%.
  Events are required to have a reconstructed vertex within 150~mm of the nominal interaction point along
  the beam axis. 
  Only events taken during stable beam conditions and satisfying detector and data-quality
requirements, which include the ID and calorimeters being in nominal operation, are
considered.
  The average number of \pp\ inelastic interactions per bunch crossing was $\mu < 1.4$.
  In \PbPb\ collisions, $\mu$ was smaller than $10^{-4}$.

\section{Jet reconstruction and analysis procedure}
\label{sec:analysis}

  The reconstruction of jets in \pp\ and \PbPb\ collisions closely follows the 
procedures described in Refs.~\cite{Aad:2012vca,Aad:2014cpa} including 
the underlying event (UE) subtraction procedure. A brief summary is given here. Jets are 
reconstructed using the anti-$k_t$ algorithm, 
which is implemented in the \verb=FastJet= software 
package~\cite{Cacciari:2011ma}. The jets are formed by 
clustering $\Delta\eta \times\Delta\phi=0.1\times\pi/32$ 
logical ``towers'' that are constructed using energy deposits in enclosed 
calorimeter cells. A background subtraction procedure based on the UE average transverse energy density, 
$\rho(\eta,\phi)$, which is calorimeter-layer dependent, was applied. 
The $\phi$ dependence is due to global azimuthal correlations between the produced particles 
(typically referred to as ``flow''). These correlations arise from the 
hydrodynamic response of the medium to the geometry of the initial collision. 
The flow contribution to the transverse energy of towers can be described by the magnitude ($v_n$) and phase ($\Psi_n$) of the 
Fourier components of the azimuthal angle distributions as:
   \begin{equation}
   \nonumber
\frac{\mbox{d}^2E_{\mbox{\tiny{T}}}}{\mbox{d}\eta\mbox{d}\phi{}}=\frac{\mbox{d}E_{\mbox{\tiny{T}}}}{\mbox{d}\eta}\left(1+2 \sum_{n}v_{n}\cos{\left(n\left(\phi-\Psi_n\right)\right)} \right),
\label{eq:flow}
\end{equation}
   where $\phi$ is the azimuthal angle of the tower and $n$ indicates the 
order of the flow harmonic.
   The modulation is dominated by $v_2$ and 
$v_3$~\cite{HION-2011-01}. 
  In this analysis, the second, third and fourth harmonics are used to further improve the UE estimation.
An iterative procedure is used to remove the effects of jets on $\rho$ 
and the $v_n$ values.
In the initial estimate of $\rho$
and $v_n$, these are estimated from the transverse energy of calorimeter cells 
within $|\eta| < 3.2$. The background is subtracted 
from calorimeter-layer-dependent transverse energies within towers associated
with the jet to obtain the subtracted jet kinematics. Then $\rho$ and 
$v_n$ values are recalculated by excluding towers within $\Delta{R} = 0.4$ 
of seed jets. Seed jets are defined as calorimeter jets with subtracted $\pt > 25$~\GeV, which are reconstructed 
with radius parameter $R=0.2$, and $R=0.4$ track jets with $\pt 
> 10$~\GeV, which are reconstructed from charged-particle tracks 
recorded in the~ID. These new $\rho$\footnote{The average $\rho$ is $\approx$270~GeV and $\approx$10~GeV in 0--10\% and 70--80\% Pb+Pb collisions, respectively.} 
and $v_n$ values are then used to evaluate a 
new subtracted energy using the original towers, and the new jet kinematic variables are calculated. 
  A final correction depending on rapidity and \pt\ is applied to obtain 
the correct hadronic energy scale for the reconstructed jets. Jets are 
calibrated using an MC-based procedure which is the same as for the ``EM+JES'' 
jets used in the analysis of \pp\ collisions~\cite{Aad:2014bia}. 
  This calibration is 
followed by a ``cross-calibration'' which relates the jet energy scale 
(JES) of \PbPb\ jets to the JES of \pp\ jets~\cite{HIjesnote}.

\begin{figure}
\centering
\includegraphics[width=0.48\textwidth]{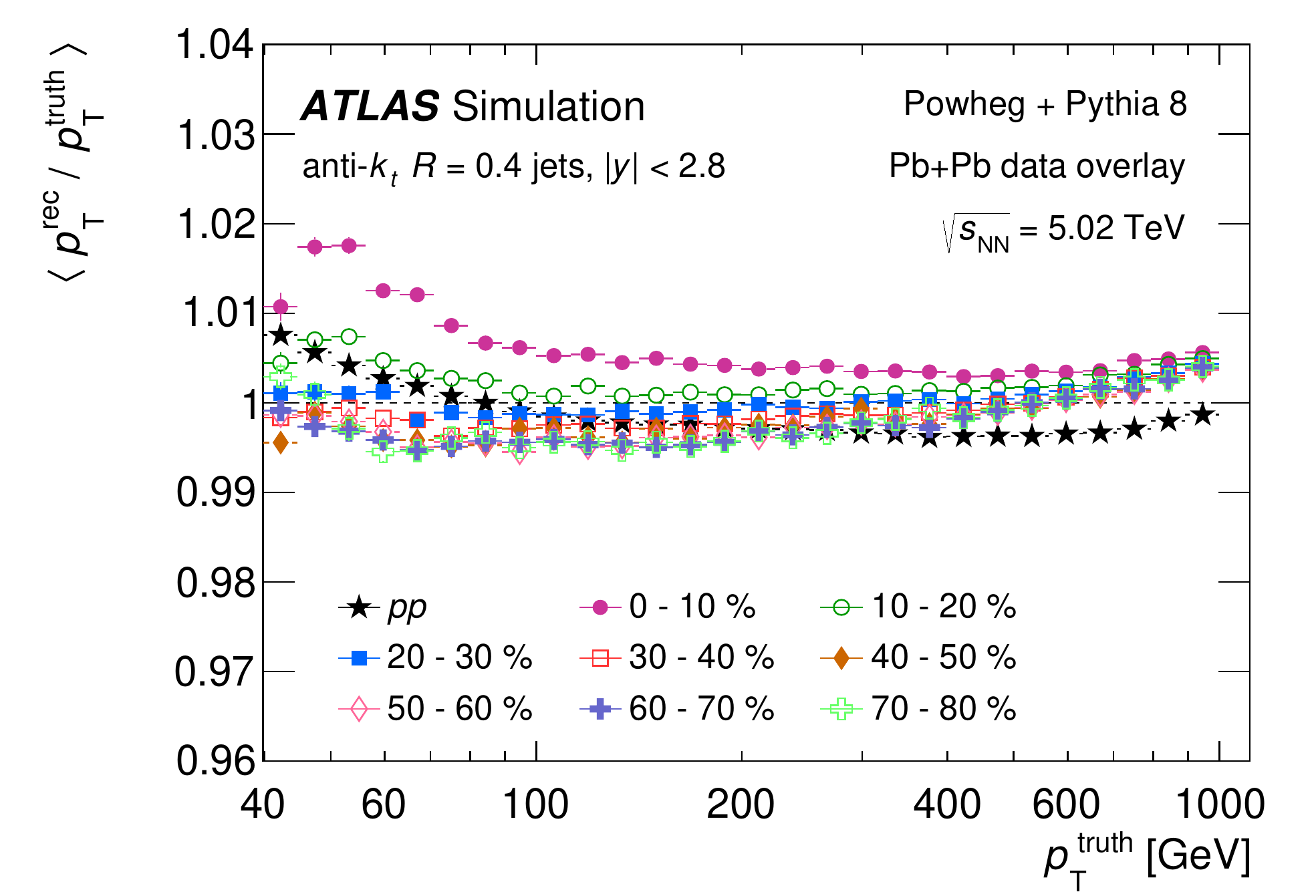}
\includegraphics[width=0.48\textwidth]{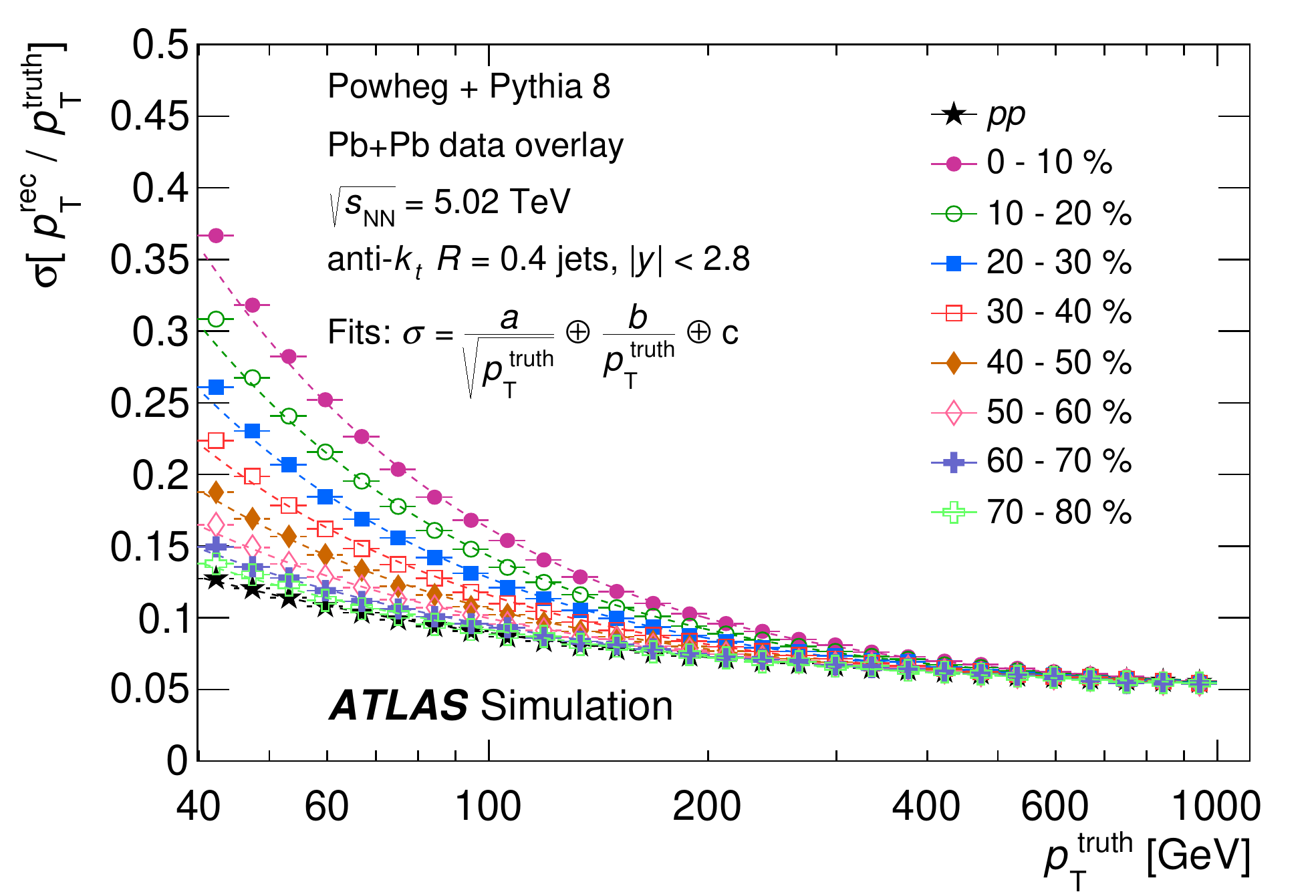}
\caption{
  The left panel shows the JES as a function of $\pt^{\mbox{\tiny{truth}}}$ and the right panel shows the JER as a function of $\pt^{\mbox{\tiny{truth}}}$ in MC 
samples. Both are for jets with $|y| < 2.8$. The curves in the right panel show fits to Eq.~(\ref{eq:jer}) for \pp, and \PbPb\ in eight centrality intervals
(0--10\%, 10--20\%, 20--30\%, 30--40\%, 40--50\%, 50--60\%, 60--70\%, and 70--80\%).
}  
\label{fig:jesjer}
\end{figure}

  The performance of the jet reconstruction was characterised by 
evaluating the JES and jet energy resolution (JER), which are correspondingly
the mean and width of the jet response 
$(\pt^{\mbox{\tiny{rec}}}/\pt^{\mbox{\tiny{truth}}})$ in the MC 
simulation. Here $\pt^{\mbox{\tiny{rec}}}$ and $\pt^{\mbox{\tiny{truth}}}$ are the transverse momenta of the reconstructed jet and truth jet, respectively. 
  The performance of the jet reconstruction in the simulation is summarised in 
Figure~\ref{fig:jesjer}, where the left and right panels show the JES 
and JER, respectively. 
The JES
  is shown as a function of $\pt^{\mbox{\tiny{truth}}}$  in the left panel of Figure~\ref{fig:jesjer}. 
It deviates from unity by less than 1\% in the kinematic region of the measurement. No rapidity dependence of the JES is observed. 
A weak centrality dependence of the JES is corrected by the unfolding procedure described later in this section.
  To express the different contributions, the JER is parameterised 
by a quadrature sum of three terms,
  \begin{equation}
  \sigma\bigg(
\frac{p_\mathrm{T}^{\mbox{\tiny{rec}}}} {p_\mathrm{T}^{\mbox{\tiny{truth}}}}
  \bigg) = \frac{a}{\sqrt{ p_\mathrm{T}^{\mbox{\tiny{truth}}}}}\oplus\frac{b}{ p_\mathrm{T}^{\mbox{\tiny{truth}}}}\oplus{c}.
  \label{eq:jer}
  \end{equation}
  The first parameter ($a$) and third parameter ($c$) 
in Eq.~(\ref{eq:jer}) are sensitive to 
the detector response and are expected to be independent of centrality, 
  while the second parameter ($b$) is centrality dependent and it is driven by UE fluctuations uncorrelated with the jet \pt. 
  The JER for different centrality 
intervals and for \pp\ collisions is shown in the right panel of 
Figure~\ref{fig:jesjer}. Fits using Eq.~(\ref{eq:jer}) are indicated with 
dashed lines. The JER is largest in the more central collisions, as 
expected from stronger fluctuations of the transverse energy in 
the UE. The JER is about 16\% for $\pt = 100$~\GeV\ in central collisions and 
decreases with increasing \pt\ to 5--6\% for jets 
with \pt\ greater than 500~\GeV. 
  The parameters $a$ and $c$ in the fit are found to be independent of centrality while the values of $b$ are consistent with the expected magnitude of UE fluctuations.
The fit parameters are listed in Table~\ref{tbl:jer} for the most central and most peripheral Pb+Pb collisions.

\begin{table}
\caption{
  The fitted parameters $a$, $b$, and $c$ (Eq.~(\ref{eq:jer})) for the most central and most peripheral collisions. 
  }
\centering
\begin{tabular}{|  r | r | r | r |} \hline
Centrality range &	$a$ [GeV$^{1/2}$] 	& 	$b$ [GeV] 	& 	$c$  			\\ \hline
70--80\%  	 &   	 0.75 $\pm$ 0.01	&	2.5 $\pm$ 0.2 	&	0.050 $\pm$ 0.001 	\\ \hline
0--10\%     	 &   	 0.76 $\pm$ 0.02 	&	14.4 $\pm$ 0.1 	&	0.049 $\pm$ 0.001	\\ \hline
\end{tabular}
\label{tbl:jer}
\end{table}

  The jet cross-section in \pp\ collisions, jet yields and \Raa\ in 
\PbPb\ collisions are measured in the following absolute rapidity ranges: 
 0--0.3, 0.3--0.8, 0.8--1.2, 1.2--1.6, 1.6--2.1, 2.1--2.8,  
 and two inclusive intervals, 0--2.1 and 0--2.8. The interval of 0--2.1 is used to make comparisons with the measurement of \Raa\ at $\sqrtsnn = 2.76$~\TeV~\cite{Aad:2014bxa}.
 The more forward region ($|y|>2.8$) is not included in the study due to deterioration of the jet reconstruction performance.  
 In \PbPb\ peripheral and \pp\ collisions, 
results are reported for 
$\pt>50$~\GeV\ and $\pt>40$~\GeV, respectively. 
  In mid-central collisions and central collisions, results are reported for 
$\pt>80$~\GeV\ and $\pt>100$~\GeV, respectively. 
  A higher value of the minimum jet \pt\ in more central \PbPb\ collisions,
compared to peripheral or \pp\ collisions, was used to reduce the contribution of jets reconstructed from fluctuations of 
the underlying events (``UE jets''). 
  These UE jets were removed by considering the charged-particle tracks with $p_\mathrm{T}^{\mbox{\tiny{trk}}}
> 4$~\GeV\ within $\Delta R=0.4$ of the jet and requiring a minimum value of $\sum p_\mathrm{T}^{\mbox{\tiny{trk}}}$. 
A threshold of $\sum p_\mathrm{T}^{\mbox{\tiny{trk}}} = 8$~\GeV\ is used throughout the 
analysis. Thresholds of $\sum p_\mathrm{T}^{\mbox{\tiny{trk}}}$ ranging from 5 to 12~\GeV\ were found to change $\Raa$ by much less than 1\% in the considered kinematic region.

  The jet \pt\ spectra are unfolded using the iterative 
  Bayesian unfolding method~\cite{DAgostini:1994zf} from the \verb=RooUnfold= software package~\cite{roounfold}, 
which accounts for bin migration due to the jet energy response.
  The response matrices used as the input to the unfolding are built 
from generator-level (truth) jets that are matched to reconstructed jets in the simulation. 
  The unmatched truth jets are incorporated as an inefficiency corrected for after the unfolding.
  In the first \pt\ bin reported in this analysis (100--126~GeV and 50--63~GeV for 0--10\% and 70--80\% Pb+Pb collisions, respectively), the relative number of unmatched truth jets is 12\% and 32\% in 0--10\% and 70--80\% collisions, respectively.
  The response matrices were generated separately for \pp\ and \PbPb\ collisions and for each rapidity and centrality interval. 
  To better represent the data, the response was reweighted along the truth-jet axis by a data-to-MC ratio.  
  The number of iterations in the unfolding was chosen so that the result is 
stable when changing the number of iterations by one. 
  Three iterations were used for \pp\ collisions while four iterations 
were used in all the centrality and rapidity intervals for \PbPb\ 
collisions.
  The unfolding procedure was tested by performing a refolding, where 
the unfolded results were convolved with the response matrix, and compared 
with the input spectra. 
  The refolded spectra were found to deviate from input spectra by less then 5\% in all centrality classes.

\section{Systematic uncertainties}
\label{sec:uncert}

\begin{figure} [t]
\centering
\includegraphics[width=0.31\textwidth]{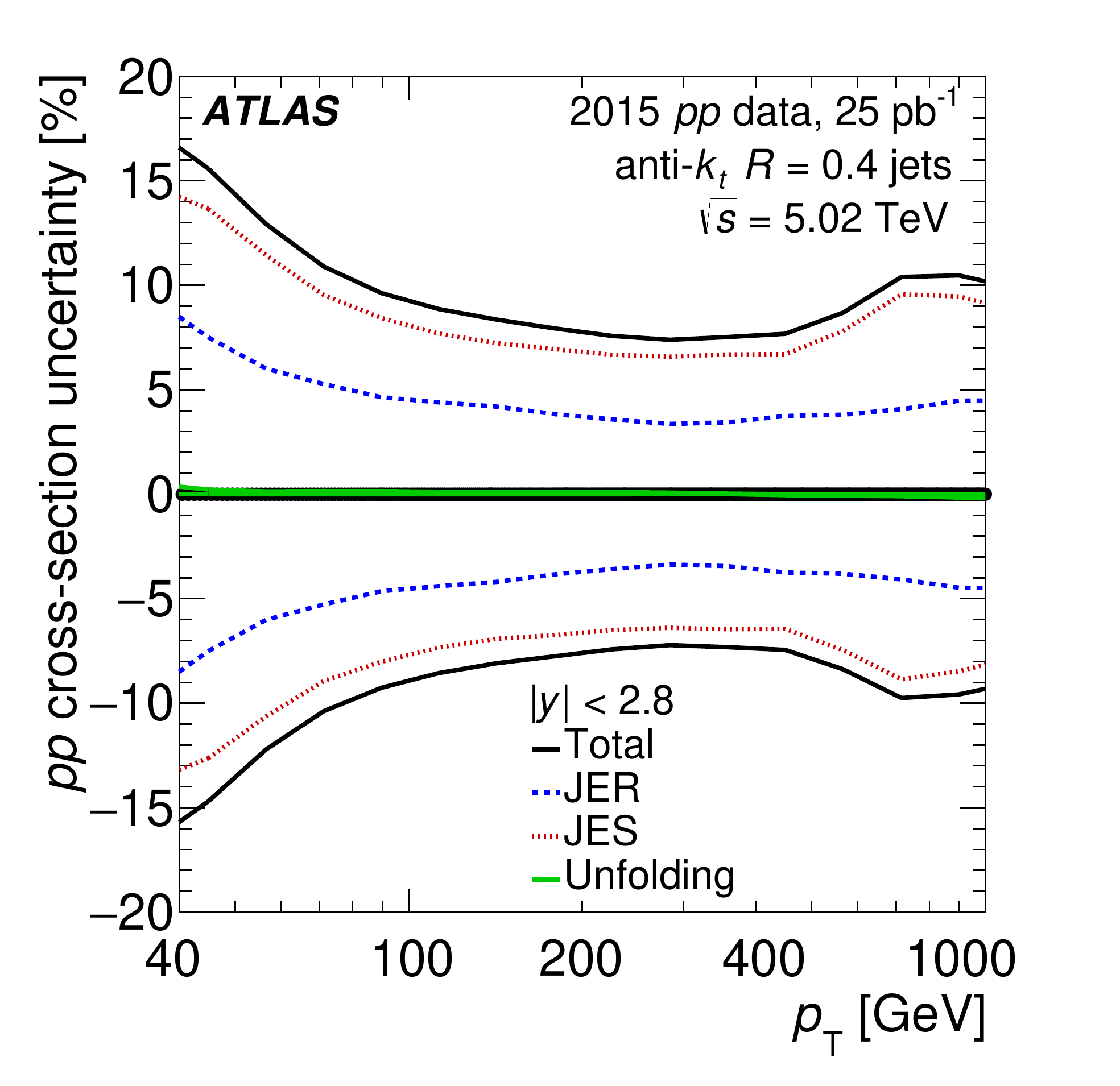}
\includegraphics[width=0.31\textwidth]{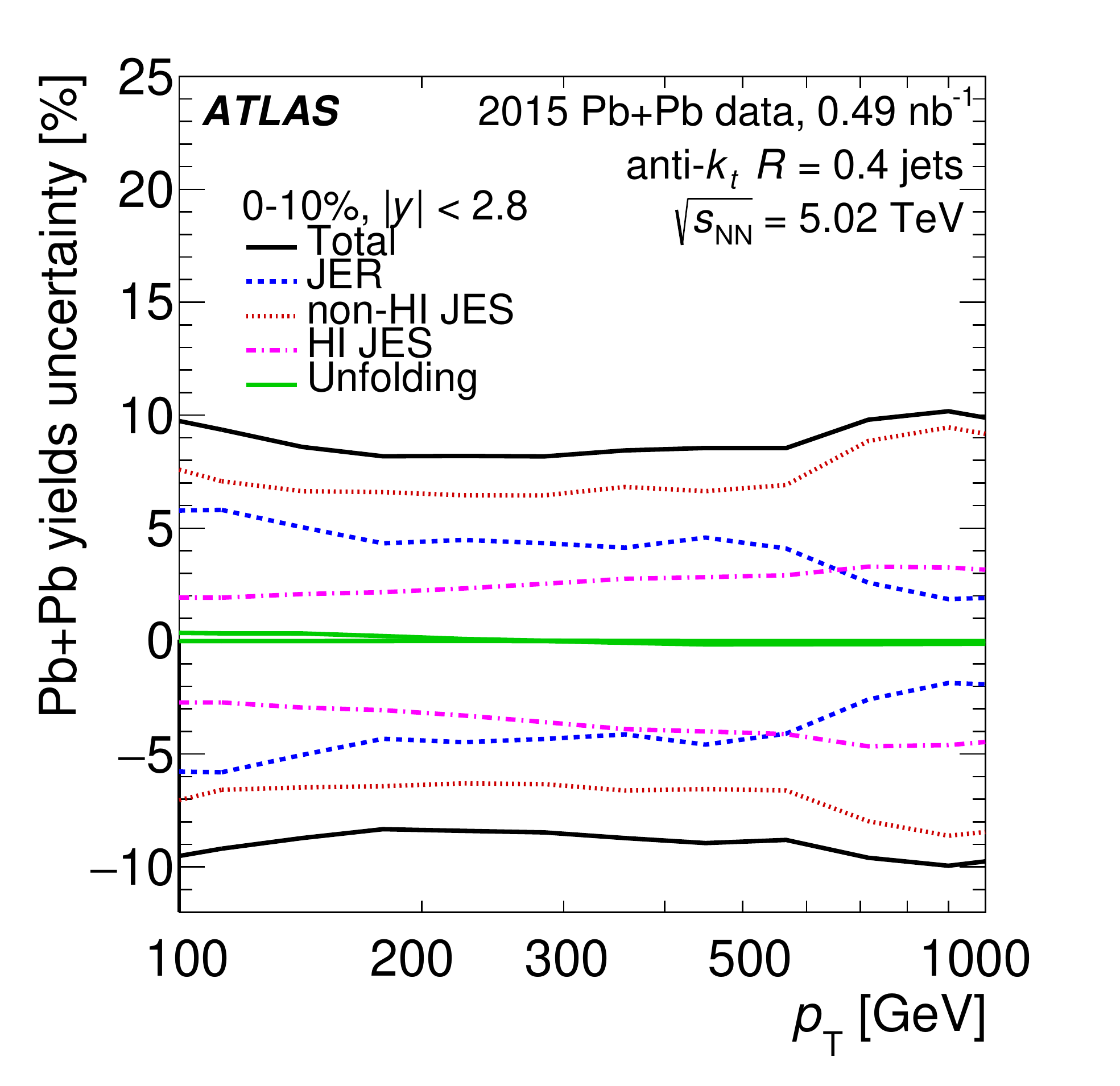}
\includegraphics[width=0.31\textwidth]{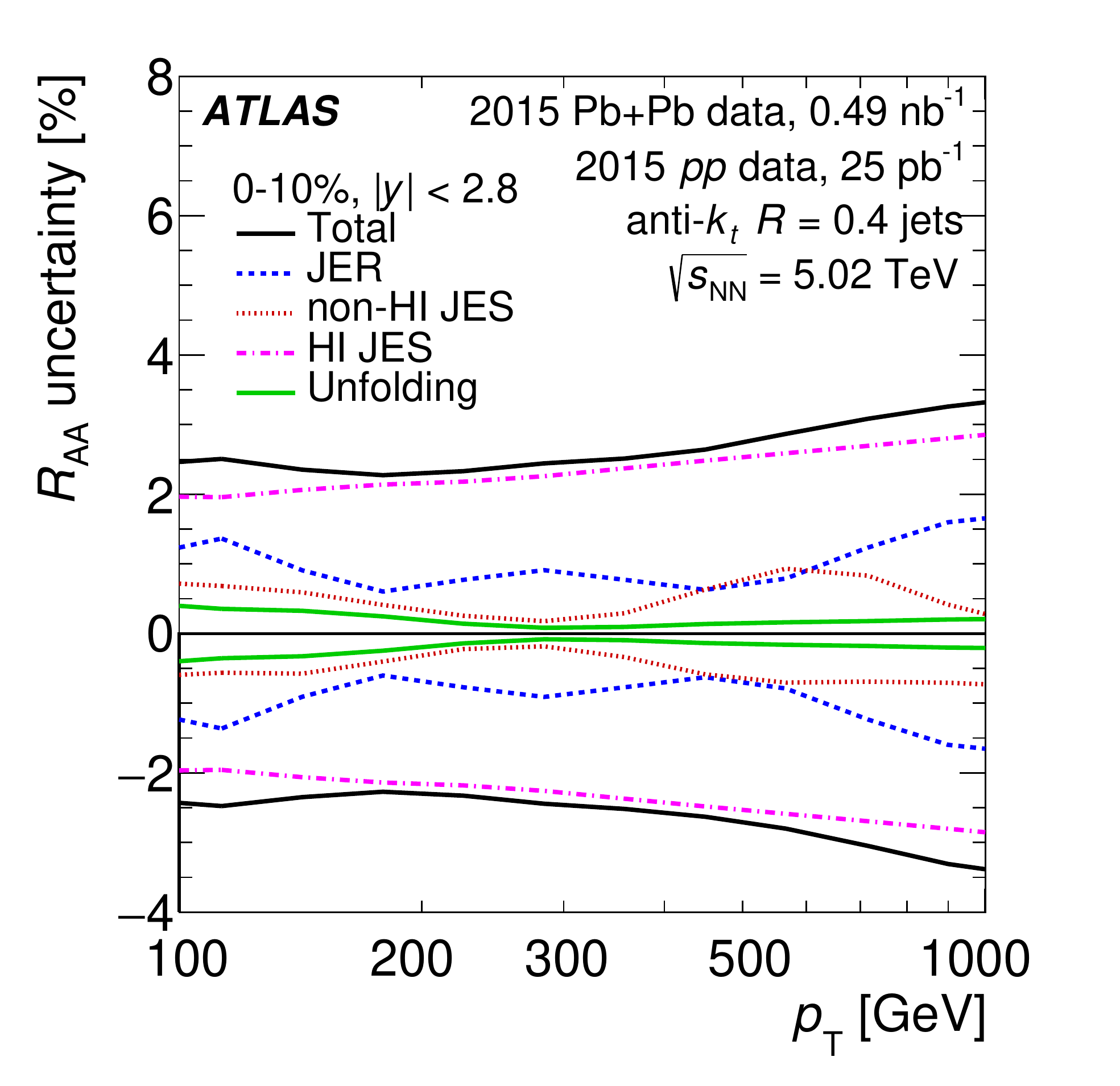}

\caption{
  Systematic uncertainties, for \pp\ jet cross-section (left), \PbPb\ jet yields (middle) and jet \Raa\ (right). Systematic uncertainties on \pp\ luminosity and \TAA, 
which affect the overall normalisation of measured distributions, are not shown.
  }
\label{fig:syst}
\end{figure}

The following sources of systematic uncertainties were identified for this 
analysis: uncertainties of the jet energy scale and jet energy resolution, uncertainty due to 
the unfolding procedure, uncertainty of the determination of the mean nuclear thickness function \TAA\ 
values, and the uncertainty of the \pp\ luminosity.
  Systematic uncertainties of the measured distributions can be 
categorised into two classes: 
bin-wise correlated uncertainties and uncertainties that affect the overall normalisation of distributions. 
Uncertainties due to the determination of \TAA\ and \pp\ luminosity belong to the second class, all other uncertainties belong to the first. 

  The strategy for determining the JES uncertainty for \PbPb\ jets is 
described in Ref.~\cite{HIjesnote}. The JES uncertainty has 
two components: the centrality-dependent component, applicable in \PbPb\ 
collisions, and a centrality-independent component, applicable in both the 
\pp\ and \PbPb\ collisions. The centrality-independent JES uncertainty 
was derived by using in situ studies of calorimeter response~\cite{ATL-PHYS-PUB-2015-015}, and 
studies of the relative 
energy scale difference between the jet reconstruction procedure in 
\PbPb\ collisions~\cite{HIjesnote} and \pp\ 
collisions~\cite{Aad:2014bia}.
  The centrality-dependent component of the JES uncertainty 
accounts for possible differences in the calorimeter response due to jets in the \PbPb\ environment. 
  It was evaluated by 
measuring the ratio of \pt\ of calorimeter jets to 
$\sum p_\mathrm{T}^{\mbox{\tiny{trk}}}$ of track jets.
 This ratio is called $\langle r_\mathrm{trk} 
\rangle$. The data-to-MC ratio of $\langle r_\mathrm{trk} \rangle$ was 
evaluated and then compared between \pp\ and \PbPb\ collisions, where it shows a 
small shift. 
  This shift may be attributed to a modification of the jet fragmentation pattern in the \PbPb\ environment which may lead to a change of the calorimeter response of jets 
reconstructed in the \PbPb\ collisions compared to jets reconstructed in \pp\ collisions. Consequently, this shift represents a typical difference in the JES between
\PbPb\ collisions and \pp\ collisions.
  It is 0.5\% in the most central 
collisions and decreases linearly to be 0\% beyond the 50--60\% centrality 
interval. This difference is taken to be the \PbPb-specific component 
of the JES uncertainty. 

Each component
that contributes to the JES uncertainty was varied separately and 
a modified response matrix was 
obtained by shifting the reconstructed jet \pt. These response 
matrices were then used to unfold the data. The difference between the 
data unfolded with the new response matrix and the nominal response 
matrix is used to determine the systematic uncertainty.

Similarly to the JES uncertainty, the systematic uncertainty due to the 
JER was obtained by performing the unfolding with 
modified response matrices. 
  The modified response matrices were generated for both the \pp\ and 
\PbPb\ collisions with the JER uncertainty which was quantified in \pp\ 
collisions using data-driven techniques~\cite{ATLAS-CONF-2015-017}. An 
additional uncertainty specific for the \PbPb\ environment 
is used, which is the uncertainty related to the impact of fluctuations 
of the UE on the JER. Both of these components are used to smear the 
reconstructed jet momentum in the MC events and regenerate the response matrices.

The results are obtained using the unfolding 
procedure with response matrices which were reweighted along the 
reconstructed jet axis to better characterise the data, as described in 
Section~\ref{sec:analysis}. The difference between the nominal results 
and results obtained with response matrices without the reweighting is 
used to calculate the uncertainty due to the unfolding procedure.

The uncertainty of the mean nuclear thickness function arises from geometric 
modelling uncertainties (e.g.\ nucleon--nucleon inelastic cross-section, 
Woods--Saxon parameterisation of the nucleon distribution~\cite{Miller:2007ri}) and the 
uncertainty of the fraction of selected inelastic Pb+Pb collisions. The 
values of these uncertainties are 
  presented in Table~\ref{tbl:npart}.

The integrated luminosity determined for 2015 \pp\ data was calibrated using data from dedicated beam separation
scans. The relative systematic uncertainty is 1.9\%, determined using procedures described in Ref.~\cite{Aad:2016ucp}.

The relative, \pt-dependent systematic uncertainties are summarised in Figure~\ref{fig:syst} for the \pp\ 
jet cross-section on the left, the \PbPb\ jet yields in the middle and the 
\Raa\ values on the right. In the \pp\ cross-section the largest uncertainty is 
from the JES, ranging from 7\% to 15\% depending on the \pt\ of the jet. The JES is 
also the largest contribution to the uncertainty in central \PbPb\ collisions 
where 
the results are reported only for jets with $\pt>100$~\GeV\ and where it
is as large as 10\%. The uncertainties of the \Raa\ values are smaller than 
those of the cross-sections and yields because the correlated 
systematic uncertainties that are common to \pp\ and \PbPb\ collisions mostly 
cancel out in the ratio. The largest contribution to the uncertainty of the 
\Raa\ values is the \PbPb\ component of the JES uncertainty, which reaches 3\% at the highest jet \pt.

\section{Results}
\label{sec:results}

\begin{figure}
\centering
\includegraphics[width=0.45\textwidth]{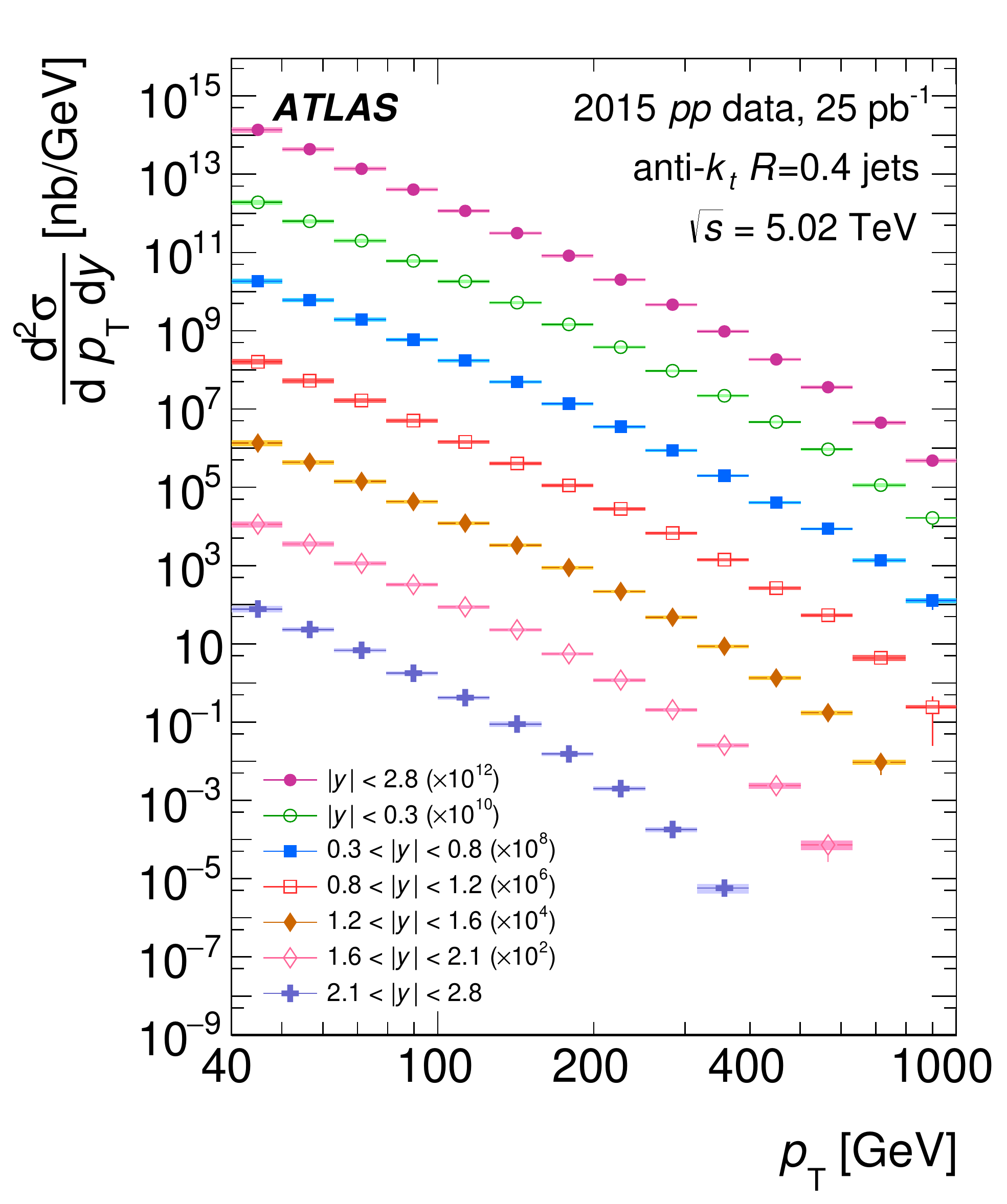}
\includegraphics[width=0.45\textwidth]{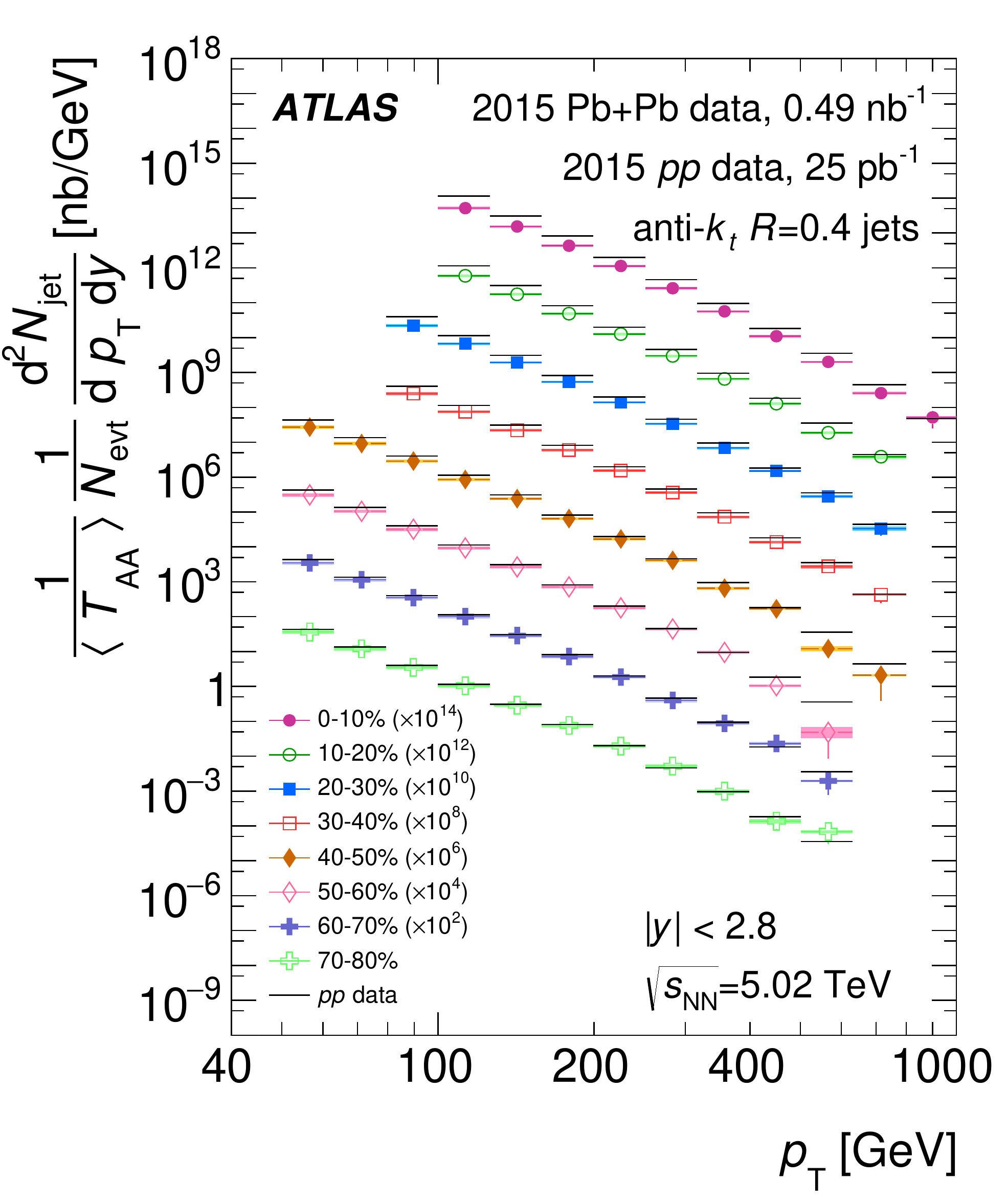}
\caption{
  Left: Inclusive jet cross-section in \pp\ collisions as a function 
of jet \pt\ in different $|y|$ intervals scaled by successive powers of $10^2$.
  Right: Per-event inclusive jet yield in \PbPb\ collisions 
normalised by \TAA\ as a function of jet \pt\ in different centrality intervals scaled by successive powers of $10^2$.
 The solid lines represent central values of \pp\ cross-section for the same 
rapidity selection scaled by the same factors to allow a comparison 
with the \PbPb\ data at different centralities.
  The error bars represent statistical uncertainties, shaded boxes 
represent systematic uncertainties.
  }
\label{fig:xsection}
\end{figure}

  The inclusive jet cross-section obtained from \pp\ collision data is 
shown in the left panel of Figure~\ref{fig:xsection}. The cross-section 
is reported for six intervals of rapidity spanning the range $|y|<2.8$
and for the whole $|y|<2.8$ interval. 
  The error bars in the figure
represent statistical uncertainties while the shaded boxes represent 
systematic uncertainties. The systematic uncertainties also include the 
uncertainty due to the luminosity, which is correlated for all the 
data points.

\begin{figure}
\centering
\includegraphics[width=0.75\textwidth]{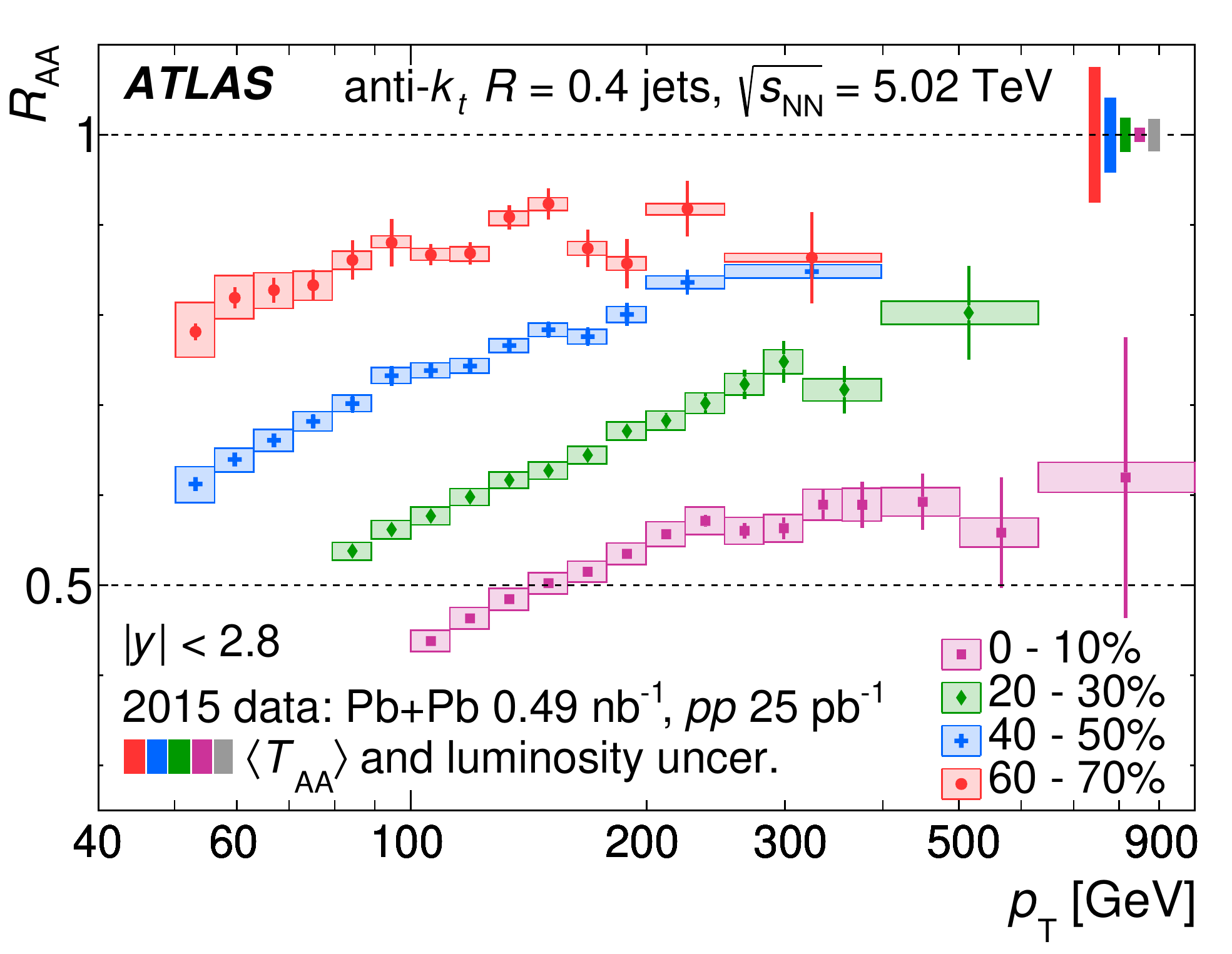} \\
\includegraphics[width=0.75\textwidth]{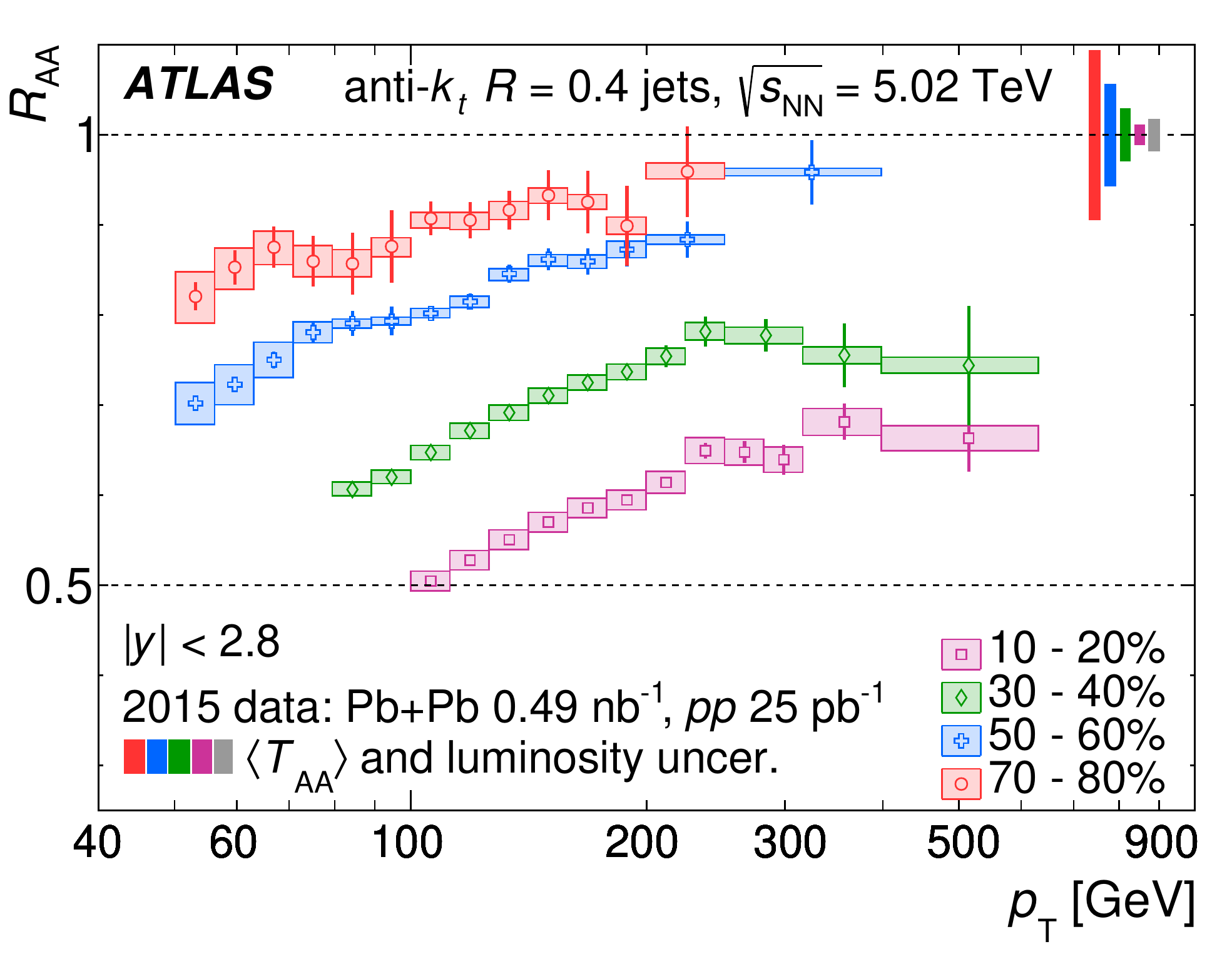} 
\caption{
  Upper panel: The \Raa\ values as a function of jet \pt\ for jets with 
$|y|<2.8$ for four centrality intervals (0--10\%, 20--30\%, 40--50\%, 60--70\%). 
  Bottom panel: The \Raa\ values as a function of jet \pt\ for jets with 
$|y|<2.8$ for four other centrality intervals (10--20\%, 30--40\%, 50--60\%, 70--80\%). 
  The error bars represent statistical uncertainties, the shaded boxes 
around the data points represent bin-wise correlated 
systematic uncertainties. 
  The coloured and grey shaded boxes 
at \Raa\ $=1$ represent fractional \TAA\ and \pp\ luminosity uncertainties, respectively, which
both affect the overall normalisation of the result.
  The horizontal size of error boxes represents the width of the \pt\ interval. 
  }
\label{fig:unfoldedRaa1}
\end{figure}

\begin{figure}
\centering
\includegraphics[width=0.94\textwidth]{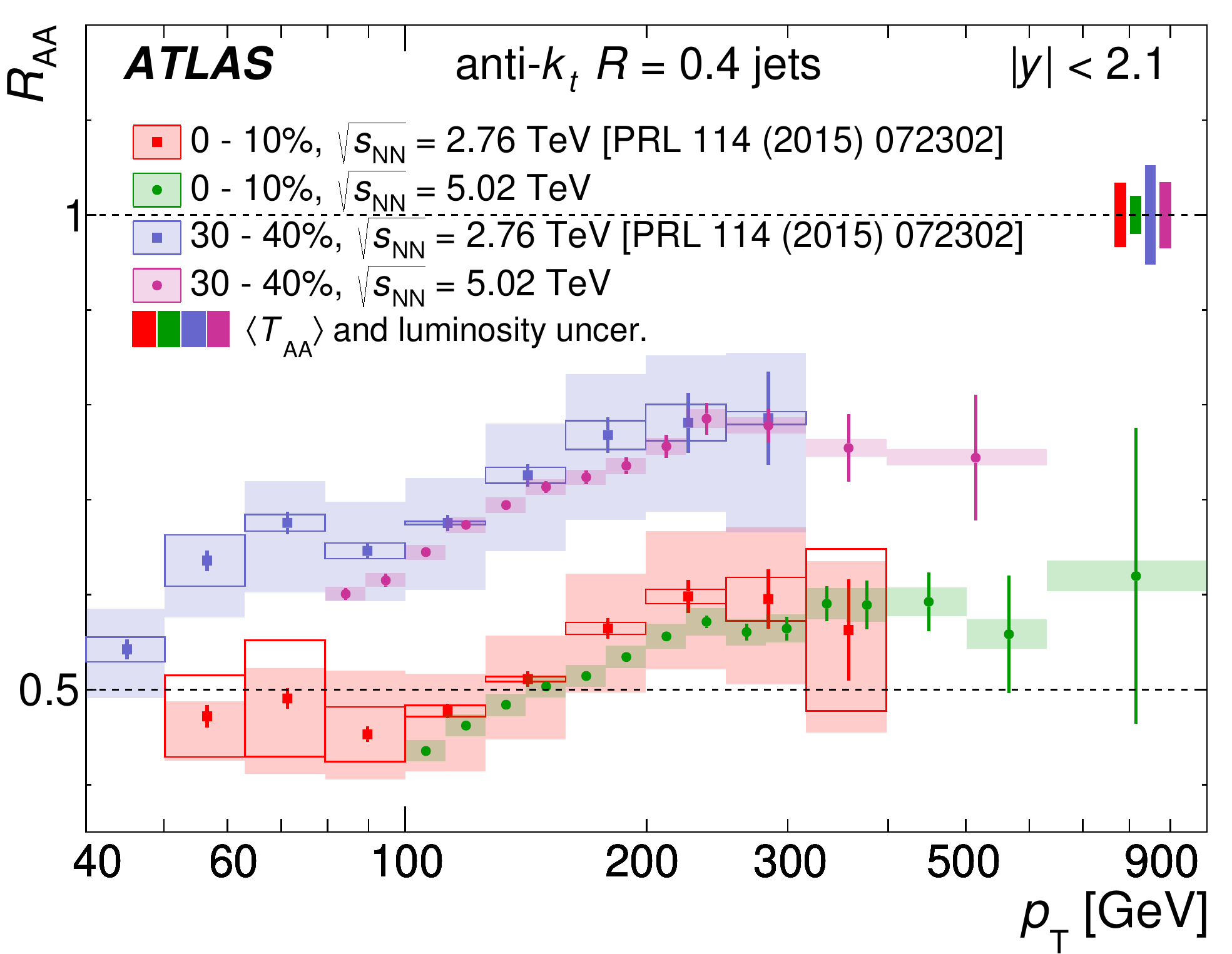}
\caption{
 The \Raa\ values as a function of jet \pt\ for jets with $|y|<2.1$ in 0--10\% and 30--40\% centrality intervals 
compared to the same quantity measured in $\sqrtsnn=2.76$~\TeV\ 
\PbPb\ collisions~\cite{Aad:2014bxa}. 
  The error bars represent statistical uncertainties, the shaded boxes 
around the data points represent bin-wise correlated 
systematic uncertainties. For $\sqrtsnn=2.76$~\TeV\ measurement, the open boxes represent uncorrelated systematic uncertainties. 
The coloured shaded boxes at \Raa\ $=1$ 
 represent the combined fractional \TAA\ and \pp\ luminosity uncertainty.
  The horizontal size of error boxes represents the width of the \pt\ interval. 
  }
\label{fig:unfoldedRaa2}
\end{figure}

The right panel of Figure~\ref{fig:xsection} shows the differential per-event \PbPb\ jet yields 
scaled by $1/\TAA$, which are presented for eight centrality intervals 
for jets with $|y|<2.8$. The solid lines  represent  the \pp\ jet  cross-sections
for the same rapidity interval; the jet yields fall below these lines, showing the jet suppression.

The nuclear modification factor evaluated as a function of jet \pt\ is 
presented in the two panels of Figure~\ref{fig:unfoldedRaa1}, 
each showing four centrality selections indicated in the legend. 
The \Raa\ value is obtained for jets with $|y|<2.8$ and with \pt\ in up to 15 intervals 
between 50 and 1000~\GeV, depending on centrality. 
 The higher \pt\ intervals are combined in the cross-section and yields 
before evaluating \Raa\ because of the large statistical 
uncertainties at high \pt.  
  A clear suppression of jet production in 
central \PbPb\ collisions relative to \pp\ collisions is observed.
  In the 0--10\% centrality interval, \Raa\ is approximately 0.45~at  
$\pt=100$~\GeV, and is observed to grow slowly (quenching decreases) with increasing jet \pt,
reaching a value of 0.6 for jets with \pt\ around 
800~\GeV. 

The \Raa\ value observed for jets with $|y|<2.1$ is compared with the 
previous measurement at $\sqrtsnn = 
2.76$~\TeV~\cite{Aad:2014bxa}. This is shown for the 
0--10\% and 30--40\% centrality intervals in Figure~\ref{fig:unfoldedRaa2}.
  The two measurements are observed to agree within their uncertainties 
in the overlapping \pt\ region.
  The apparent reduction of the size of systematic uncertainties in the new measurement is driven by collecting the \pp\ and \PbPb\ data during the same LHC running period.

\begin{figure}
\centering
\includegraphics[width=0.63\textwidth]{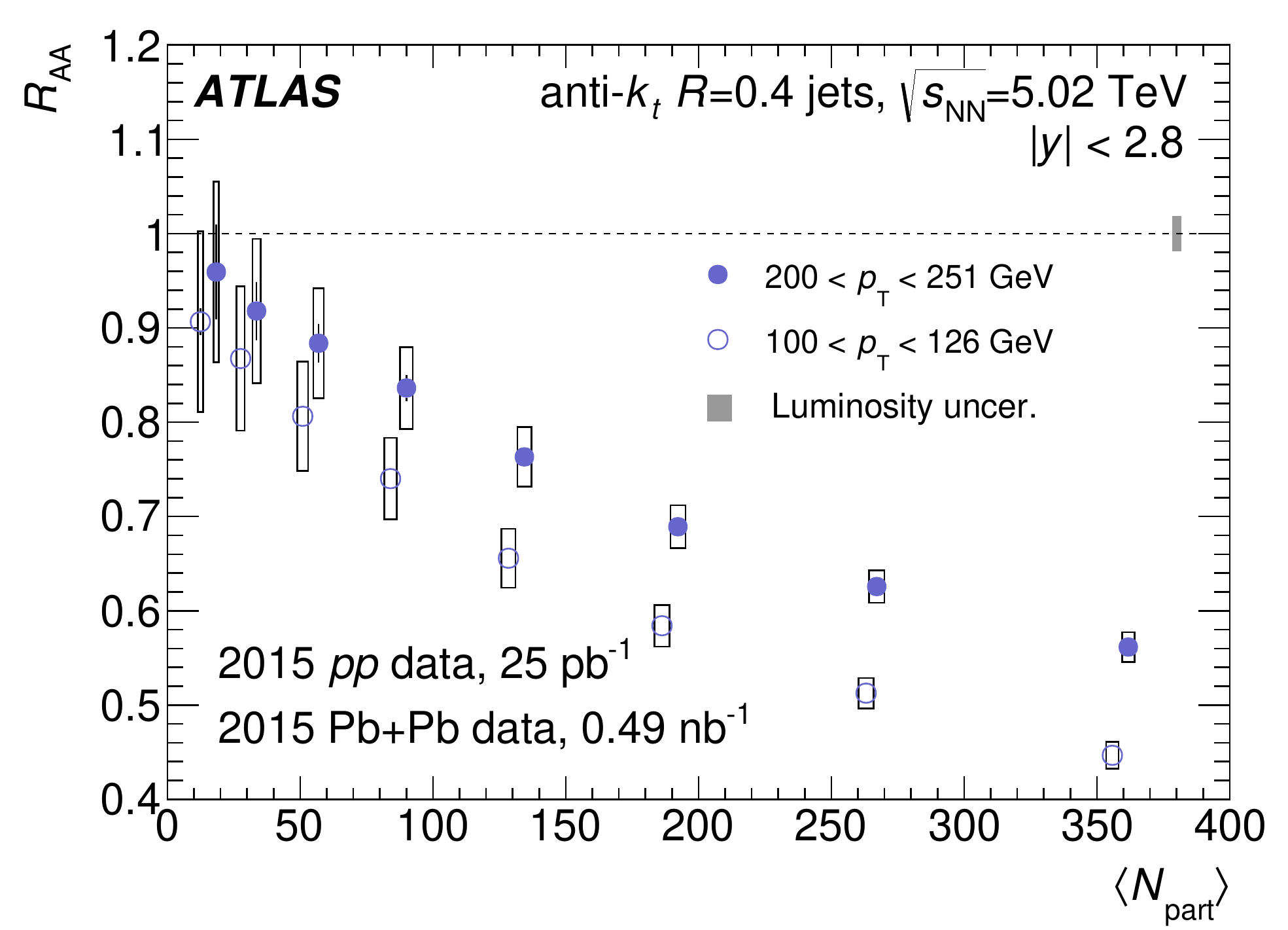}
\caption{
  The \Raa\ values for jets with $100 < \pt < 126$~\GeV\ and $200 < \pt <251$~\GeV\ for rapidity $|y|<2.8$ evaluated as a function of $\ANpart$. For legibility, the 
$\ANpart$ values are shifted by $-7$ and $+7$ for $100 < \pt < 126$~\GeV\ selection and $200 < \pt <251$~\GeV\ selection, respectively.
  The error bars represent statistical uncertainties. The heights of the open boxes represent systematic uncertainties. 
  The widths of the open boxes represent the uncertainties in the determination of
$\ANpart$. 
  The grey shaded box at unity represents the uncertainty of the \pp\ integrated
luminosity.
  }
\label{fig:unfoldedRaa3}
\end{figure}

  The \ANpart\ dependence of \Raa\ is shown in 
Figure~\ref{fig:unfoldedRaa3} for jets with $|y|<2.8$ and
for two representative \pt\ intervals: $100 < \pt < 126$~\GeV\ and $200 < \pt <251$~\GeV.
  The open boxes around the data points represent the bin-wise 
correlated systematic uncertainties which include also the uncertainty 
of \TAA. 
  A smooth evolution of \Raa\ is observed, with the 
largest values of \Raa\ in the most peripheral collisions and the 
smallest values of \Raa\ in the most central collisions. 
  The magnitude of \Raa\ is observed to be larger        
for jets in higher \pt\ interval for $\ANpart \gtrsim 50$. For $\ANpart \lesssim 50$ the difference is not statistically significant.

\begin{figure}
\centering
\includegraphics[width=0.8\textwidth]{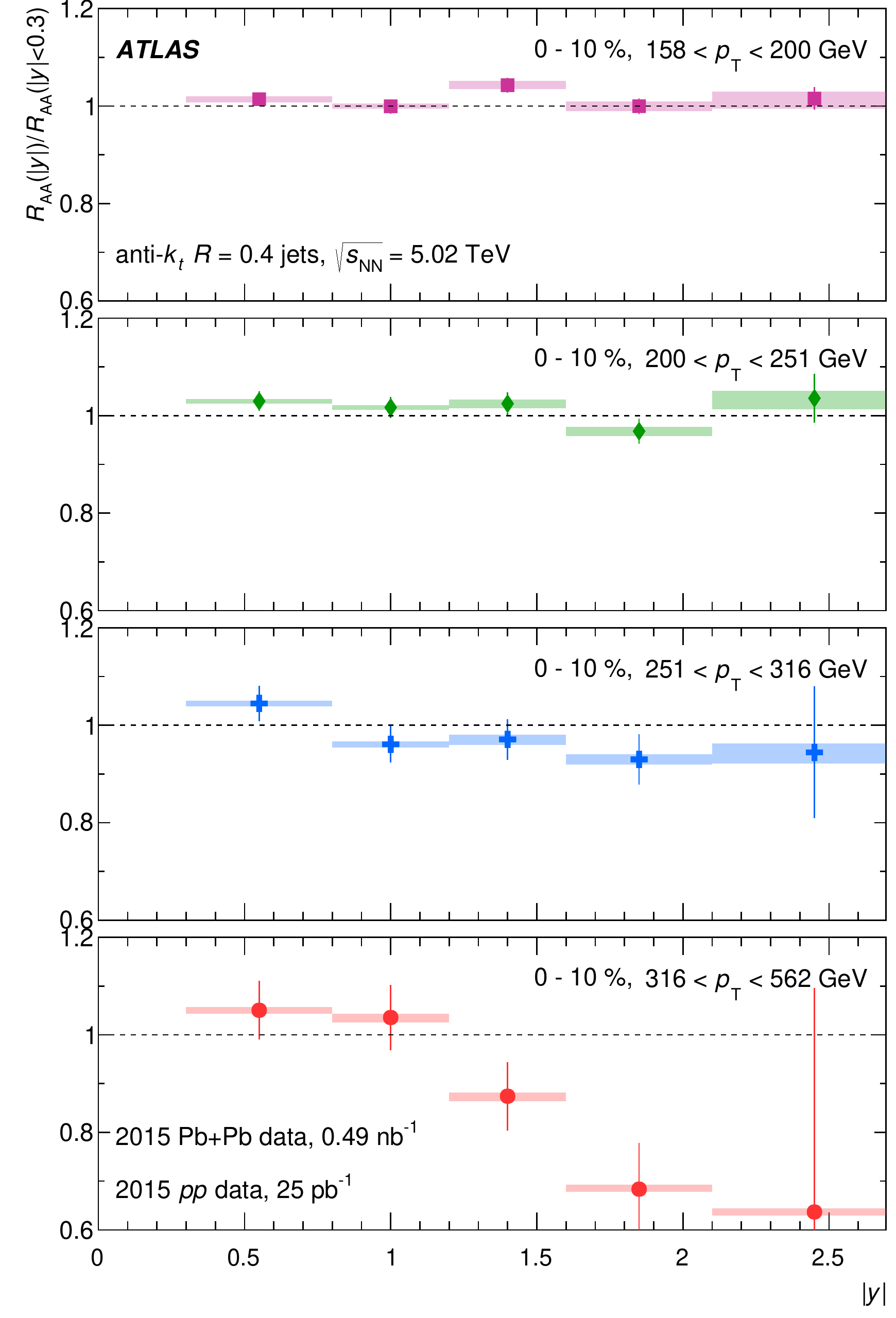}
\caption{
 The ratio of \Raa\ to the \Raa\ value for $|y| < 0.3$ as a function of $|y|$ for jets in four \pt\ intervals ($158 < \pt < 200$ \GeV, 200 $< \pt <$ 
251 \GeV, 251 $< \pt <$ 316 \GeV, and 316 $< \pt <$ 562 \GeV) shown for the 10\% most central \PbPb\ collisions.  
The error bars represent statistical uncertainties, the shaded boxes around the data points represent systematic uncertainties.
 }
\label{fig:unfoldedRaa4}
\end{figure}

The rapidity dependence of \Raa\ is shown in 
Figure~\ref{fig:unfoldedRaa4} as the ratio of \Raa\ to its value 
measured for $|y| < 0.3$. This representation 
was chosen because all systematic uncertainties largely cancel out in the 
ratio. The distributions are reported in intervals of increasing values of \pt\ in the 
four panels. The ratio is constant in rapidity 
at lower \pt. As the \pt\ increases, the value of \Raa\ starts to decrease 
with rapidity and the decrease is most significant in the highest 
\pt\ interval of 316--562~GeV. In this \pt\ interval, the value of the \Raa\ ratio 
is $0.83 \pm 0.07$ and $0.68 \pm 0.13$ in the rapidity regions of $|y| = 1.2$--2.8 and $|y| = 1.6$--2.8, respectively.
This decrease was predicted in Ref.~\cite{Spousta:2015fca} as a consequence of a steepening of jet \pt\ spectra in the forward rapidity region.

\begin{figure}
\centering
\includegraphics[width=0.9\textwidth]{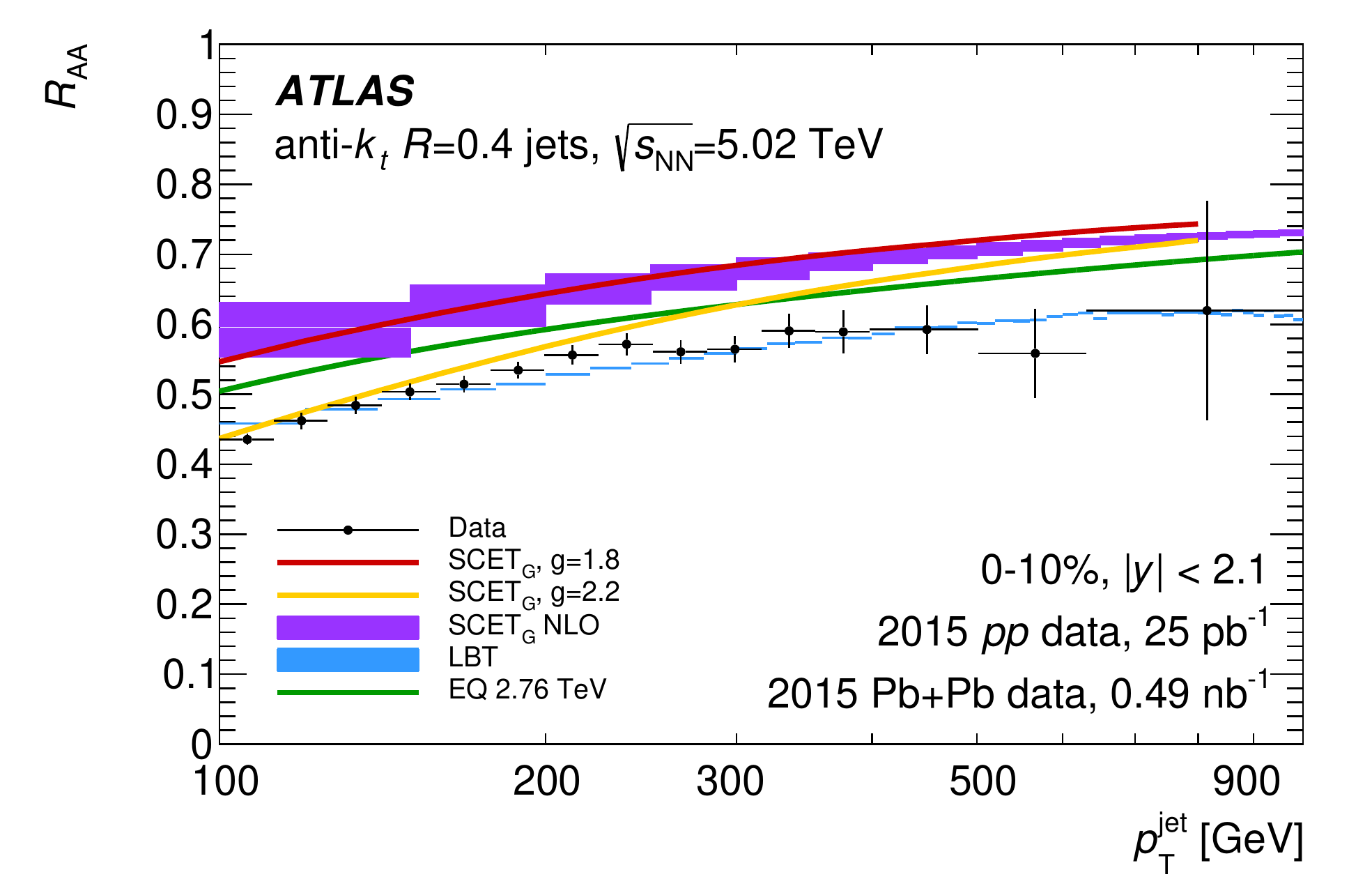}
\caption{
  The \Raa\ values as a function of jet \pt\ for the 0--10\% centrality interval and 
$|y|<2.1$ compared with theory predictions. The uncertainties 
of the data points are the combined statistical and systematic 
uncertainties. The vertical width of the distribution shown for the LBT and SCET$_\mathrm{G}$ NLO models represents the uncertainty of the theory 
prediction.
 }
\label{fig:compareWithTheory}
\end{figure}

A comparison of the \Raa\ values with theoretical predictions is provided in 
Figure~\ref{fig:compareWithTheory}. The \Raa\ values obtained as a function of jet \pt\  are compared 
with five predictions for jets with $|y|<2.1$ where theory calculations are available: the Linear Boltzmann Transport model (LBT) 
\cite{He:2015pra}, three calculations using the Soft Collinear Effective 
Theory approach (SCET$_\mathrm{G}$) 
\cite{Chien:2015vja,Chien:2015hda,Vitev:2008rz,Kang:2017frl}, and the 
Effective Quenching model (EQ) \cite{Spousta:2015fca}.
  The LBT model combines a kinetic description of parton propagation with a
hydrodynamic description of the underlying medium evolution while 
keeping track of thermal recoil partons from each scattering and 
their further propagation in the medium \cite{He:2015pra}. The
  SCET$_\mathrm{G}$ approach uses semi-inclusive jet functions~\cite{Kang:2016mcy} evaluated with 
in-medium parton splittings computed using soft collinear effective 
theory. It provides three predictions with two different settings of the strong 
coupling constant associated with the jet--medium interaction ($g=2.2$ 
and $g=1.8$) and the calculation at NLO accuracy.
  The EQ model incorporates energy loss effects through two downward shifts in the
\pt\ spectrum based on a semi-empirical parameterisation of jet quenching effects. One shift is applied to quark-initiated jets and a larger shift to 
gluon-initiated jets. 
The EQ model requires experimental data in order to extract the parameters of the energy loss.
The same parameters of the jet energy loss as for 
$\sqrtsnn = 2.76$~\TeV\ data \cite{Spousta:2015fca} are used here. 
  All the models are capable of reproducing the general trends seen in the
data. For $\pt \lesssim 250$~\GeV, the data agrees best with the SCET$_\mathrm{G}$ model which uses $g=2.2$. For $\pt \gtrsim 250$~\GeV\ the LBT model describes the data better.
Disagreement between the data and the EQ model using the parameters of
the jet energy loss from 2.76~\TeV~\PbPb\ data can be explained as a
consequence of stronger quenching in 5.02~\TeV~\PbPb\ collisions.

\clearpage
\newpage

\section{Summary}
\label{sec:summary}

Measurements of inclusive jet yields in \PbPb\ collisions, jet cross-sections in \pp\ collisions, and the jet nuclear modification factor, 
\Raa, are performed using 0.49 nb$^{-1}$ of \PbPb\ collision data
and 25 pb$^{-1}$ of \pp\ collision data collected at the same nucleon--nucleon centre-of-mass 
energy of 5.02~\TeV\ by the ATLAS detector at the 
LHC. Jets, reconstructed using the \antikt\ algorithm with radius parameter $R=0.4$, are 
measured over the transverse momentum range of 40--1000~\GeV\ in six 
rapidity intervals covering $|y|<2.8$. 
The jet yields measured in \PbPb\ collisions are suppressed
relative to the jet cross-section measured in \pp\ collisions scaled by the mean nuclear thickness function, \TAA.
  The magnitude of \Raa\ increases with increasing jet transverse momentum, reaching a value of approximately~0.6
  at 1~\TeV\ in the most central collisions. The magnitude of \Raa\ also increases
towards peripheral collisions.
  The \Raa\ value is 
independent of rapidity at low jet \pt. For jets with \pt~$\gtrsim 300$~\GeV\ a sign of a decrease with rapidity is observed.
  The magnitude of the jet suppression as well as its evolution with jet 
\pt\ and rapidity are consistent with those reported in a similar 
measurement performed with \PbPb\ collisions at $\sqrtsnn = 2.76$~\TeV\ in 
the kinematic region where the two measurements overlap.

The results presented here extend previous measurements 
to significantly higher transverse momenta and larger rapidities of jets 
and improve on the precision of the measurement. This allows 
precise and detailed comparisons of the data to theoretical models of 
jet quenching. These new results can also be used as additional input 
to understand the centre-of-mass energy dependence of jet suppression.

\section*{Acknowledgements}

We thank CERN for the very successful operation of the LHC, as well as the
support staff from our institutions without whom ATLAS could not be
operated efficiently.

We acknowledge the support of ANPCyT, Argentina; YerPhI, Armenia; ARC, Australia; BMWFW and FWF, Austria; ANAS, Azerbaijan; SSTC, Belarus; CNPq and FAPESP, Brazil; NSERC, NRC and CFI, Canada; CERN; CONICYT, Chile; CAS, MOST and NSFC, China; COLCIENCIAS, Colombia; MSMT CR, MPO CR and VSC CR, Czech Republic; DNRF and DNSRC, Denmark; IN2P3-CNRS, CEA-DRF/IRFU, France; SRNSFG, Georgia; BMBF, HGF, and MPG, Germany; GSRT, Greece; RGC, Hong Kong SAR, China; ISF and Benoziyo Center, Israel; INFN, Italy; MEXT and JSPS, Japan; CNRST, Morocco; NWO, Netherlands; RCN, Norway; MNiSW and NCN, Poland; FCT, Portugal; MNE/IFA, Romania; MES of Russia and NRC KI, Russian Federation; JINR; MESTD, Serbia; MSSR, Slovakia; ARRS and MIZ\v{S}, Slovenia; DST/NRF, South Africa; MINECO, Spain; SRC and Wallenberg Foundation, Sweden; SERI, SNSF and Cantons of Bern and Geneva, Switzerland; MOST, Taiwan; TAEK, Turkey; STFC, United Kingdom; DOE and NSF, United States of America. In addition, individual groups and members have received support from BCKDF, the Canada Council, CANARIE, CRC, Compute Canada, FQRNT, and the Ontario Innovation Trust, Canada; EPLANET, ERC, ERDF, FP7, Horizon 2020 and Marie Sk{\l}odowska-Curie Actions, European Union; Investissements d'Avenir Labex and Idex, ANR, R{\'e}gion Auvergne and Fondation Partager le Savoir, France; DFG and AvH Foundation, Germany; Herakleitos, Thales and Aristeia programmes co-financed by EU-ESF and the Greek NSRF; BSF, GIF and Minerva, Israel; BRF, Norway; CERCA Programme Generalitat de Catalunya, Generalitat Valenciana, Spain; the Royal Society and Leverhulme Trust, United Kingdom.

The crucial computing support from all WLCG partners is acknowledged gratefully, in particular from CERN, the ATLAS Tier-1 facilities at TRIUMF (Canada), NDGF (Denmark, Norway, Sweden), CC-IN2P3 (France), KIT/GridKA (Germany), INFN-CNAF (Italy), NL-T1 (Netherlands), PIC (Spain), ASGC (Taiwan), RAL (UK) and BNL (USA), the Tier-2 facilities worldwide and large non-WLCG resource providers. Major contributors of computing resources are listed in Ref.~\cite{ATL-GEN-PUB-2016-002}.

\printbibliography

\clearpage
\input{atlas_authlist}

\end{document}

%% file: HION-2017-10-PAPER-metadata.tex
\AtlasTitle{Measurement of the nuclear modification factor for inclusive jets in Pb+Pb collisions at $\sqrt{s_\mathrm{NN}}=5.02$~\TeV\ with the ATLAS detector}

\AtlasJournal{Phys.\ Lett.\ B.}

\AtlasAbstract{
  Measurements of the yield and nuclear modification factor, $\Raa$, for inclusive jet production are performed using $0.49$~nb$^{-1}$ of \PbPb\ data at 
$\sqrt{s_\mathrm{NN}} = 5.02$~\TeV\ and 25~pb$^{-1}$ of \pp\ data at $\sqrt{s}=5.02$~\TeV\ with the ATLAS detector at the LHC. Jets are reconstructed with the \antikt\ 
algorithm with radius parameter $R=0.4$ and are measured over the transverse momentum range of 40--1000~\GeV\ in six rapidity intervals covering $|y|<2.8$. The 
magnitude of \Raa\ increases with increasing jet transverse momentum, reaching a value of approximately~0.6 at 1~\TeV\ in the most central collisions. The magnitude of 
\Raa\ also increases towards peripheral collisions. The value of \Raa\ is independent of rapidity at low jet transverse momenta, but it is observed to decrease with 
increasing rapidity at high transverse momenta.
  }

\author{The ATLAS Collaboration}

\AtlasRefCode{HION-2017-10}

\PreprintIdNumber{CERN-EP-2018-105}

%% file: atlas_authlist.tex
 
\begin{flushleft}
{\Large The ATLAS Collaboration}

\bigskip

M.~Aaboud$^\textrm{\scriptsize 34d}$,    
G.~Aad$^\textrm{\scriptsize 99}$,    
B.~Abbott$^\textrm{\scriptsize 124}$,    
O.~Abdinov$^\textrm{\scriptsize 13,*}$,    
B.~Abeloos$^\textrm{\scriptsize 128}$,    
D.K.~Abhayasinghe$^\textrm{\scriptsize 91}$,    
S.H.~Abidi$^\textrm{\scriptsize 164}$,    
O.S.~AbouZeid$^\textrm{\scriptsize 39}$,    
N.L.~Abraham$^\textrm{\scriptsize 153}$,    
H.~Abramowicz$^\textrm{\scriptsize 158}$
% ...
\bigskip
\\

% ...
$^{13}$Institute of Physics, Azerbaijan Academy of Sciences, Baku; Azerbaijan.\\
% ...
$^{34}$$^{(a)}$Facult\'e des Sciences Ain Chock, R\'eseau Universitaire de Physique des Hautes Energies - Universit\'e Hassan II, Casablanca;$^{(b)}$Centre National de l'Energie des Sciences Techniques Nucleaires (CNESTEN), Rabat;$^{(c)}$Facult\'e des Sciences Semlalia, Universit\'e Cadi Ayyad, LPHEA-Marrakech;$^{(d)}$Facult\'e des Sciences, Universit\'e Mohamed Premier and LPTPM, Oujda;$^{(e)}$Facult\'e des sciences, Universit\'e Mohammed V, Rabat; Morocco.\\
% ...
$^{39}$Niels Bohr Institute, University of Copenhagen, Copenhagen; Denmark.\\
% ...
$^{91}$Department of Physics, Royal Holloway University of London, Egham; United Kingdom.\\
% ...
$^{99}$CPPM, Aix-Marseille Universit\'e, CNRS/IN2P3, Marseille; France.\\
% ...
$^{124}$Homer L. Dodge Department of Physics and Astronomy, University of Oklahoma, Norman OK; United States of America.\\
% ...
$^{128}$LAL, Universit\'e Paris-Sud, CNRS/IN2P3, Universit\'e Paris-Saclay, Orsay; France.\\
% ...
$^{153}$Department of Physics and Astronomy, University of Sussex, Brighton; United Kingdom.\\
% ...
$^{158}$Raymond and Beverly Sackler School of Physics and Astronomy, Tel Aviv University, Tel Aviv; Israel.\\
% ...
$^{164}$Department of Physics, University of Toronto, Toronto ON; Canada.\\
% ...
$^{a}$ Also at Borough of Manhattan Community College, City University of New York, NY; United States of America.\\
$^{b}$ Also at California State University, East Bay; United States of America.\\
$^{c}$ Also at Centre for High Performance Computing, CSIR Campus, Rosebank, Cape Town; South Africa.\\
$^{d}$ Also at CERN, Geneva; Switzerland.\\
$^{e}$ Also at CPPM, Aix-Marseille Universit\'e, CNRS/IN2P3, Marseille; France.\\
% ...
$^{*}$ Deceased

\end{flushleft}
